\documentclass[aps,prl,twocolumn]{revtex4}
\usepackage{graphicx}
\usepackage{amsmath}
\usepackage{array}
\usepackage{mathptmx}
\usepackage{anyfontsize}
\usepackage{t1enc}
\begin{document}

\title {
Strong Dynamical Heterogeneity and Universal Scaling in Driven
Granular Fluids} 

\author{Karina E. Avila$^{1,2}$} 
\email[]{karina.avila@theorie.physik.uni-goettingen.de}
\author{Horacio E. Castillo$^{1}$}
\author{Andrea Fiege$^{3}$}
\author{Katharina Vollmayr-Lee$^{4}$}
\author{Annette Zippelius$^{3,2}$}

\affiliation{$^{1}$Department of Physics and Astronomy and Nanoscale
  and Quantum Phenomena Institute, Ohio University, Athens, Ohio 45701,
  USA\\ 
$^{2}$Max-Planck-Institut f\"ur Dynamik und Selbstorganisation, Am Fassberg 17, D-37077 G\"ottingen, Germany\\
$^{3}$Institut f\"ur Theoretische Physik, Georg-August-Universit\"at G\"ottingen, Friedrich-Hund-Platz 1, D-37077 G\"ottingen, Germany\\
$^{4}$ Department of Physics and Astronomy, Bucknell University, Lewisburg, Pennsylvania 17837, USA}

\date{\today}

\begin{abstract}
  Large scale simulations of two-dimensional bidisperse granular
  fluids allow us to determine spatial correlations of slow particles
  via the four-point structure factor $S_4(q,t)$. Both cases,
  elastic ($\varepsilon=1$) as well as inelastic ($\varepsilon < 1$)
  collisions, are studied. As the fluid approaches structural arrest,
  i.e. for packing fractions in the range $0.6 \le \phi \le 0.805$,
  scaling is shown to hold: $S_4(q,t)/\chi_4(t)=s(q\xi(t))$. Both the
  dynamic susceptibility, $\chi_4(\tau_{\alpha})$, as well as the
  dynamic correlation length, $\xi(\tau_{\alpha})$, evaluated at the
  $\alpha$ relaxation time, $\tau_{\alpha}$, can be fitted to a power
  law divergence at a critical packing fraction. The measured
  $\xi(\tau_{\alpha})$ widely exceeds the largest one previously
  observed for hard sphere 3d fluids. 
  The number of particles in a slow cluster and the correlation length
  are related by a robust power law, $\chi_4(\tau_{\alpha}) \approx
  \xi^{d-p}(\tau_{\alpha})$, with an exponent $d-p\approx 1.6$. This
  scaling is remarkably independent of $\varepsilon$, even though the
  strength of the dynamical heterogeneity increases dramatically as
  $\varepsilon$ grows.
\end{abstract}

%
%
%

\maketitle

Viscous liquids, colloidal suspensions, and granular fluids are all
capable of undergoing dynamical arrest, either by reducing the
temperature in the case of viscous liquids, or by increasing the
density in the cases of colloidal suspensions and of granular
systems~\cite{Debenedetti2001, Pusey1986, OHern2003, Berthier2011}. As
the dynamical arrest is approached, not only does the dynamics become
dramatically slower, but it becomes increasingly
heterogeneous~\cite{Berthier2011, Ediger2000, Sillescu, Russell2000,
  Kegel2000, Weeks2000, Cipelleti, Doliwa2000, Lacevic2003, Abate2007,
  Keys2007, Toninelli2005, Flenner2011, Dauchot2005, Marty2005,
  Candelier2009, Lechenault2008a, Lechenault2008b}. One of the most
common ways to characterize the heterogeneity in the dynamics is to
probe its fluctuations~\cite{Berthier2011}. Since probing the dynamics
requires observing the system at two times, probing the spatial
fluctuations in the dynamics naturally leads to defining quantities
that correlate the changes in the state of the system between two
times, at two spatial points, i.e. four-point functions.  Those
quantities include the dynamic susceptibility $\chi_4(t)$, which gives
a spatially integrated measurement of the total fluctuations, and the
four point structure factor $S_4(q,t)$, which is the Fourier transform
of the spatial correlation function describing the local fluctuations
in the dynamics~\cite{Berthier2011, Lacevic2003, Toninelli2005}. From
the small wave-vector behavior of $S_4(q,t)$, a correlation length
$\xi(t)$ can be extracted, and it has been found in simulations of
viscous liquids and dense colloidal suspensions that this correlation
length grows as dynamical arrest is approached~\cite{Berthier2011,
  Lacevic2003, Toninelli2005, Flenner2011}. For granular matter, on
the other hand, the jamming transition has been analyzed extensively,
but studies on dynamic heterogeneity (DH) are few.  Two experimental
groups have investigated driven $2d$ granular beds in the context of
DH. These studies are restricted to small systems of order a few
thousand particles~\cite{Abate2007, Keys2007, Dauchot2005, Marty2005,
  Candelier2009,Lechenault2008a, Lechenault2008b}.  $\chi_4(t)$ has
been measured, but spatial correlations have not been investigated
systematically due to small system size. Instead, compact regions of
correlated particles are usually assumed, $\chi_4(t)\sim \xi^d(t)$,
thereby determining a correlation length $\xi(t)$.

Here we determine $\xi$ and $\chi_4$ {\it independently} from
$S_4(q,t)$ -- without further assumption. We show that there is indeed
a cooperative lengthscale which grows dramatically as structural
arrest is approached: Varying the density by $10\%$ results in an
increase in $\xi$ by a factor $\sim 20$. The number of correlated
particles is given by $\chi_4(t)$, which increases by a factor $>10^2$
in the same range of densities. Both $\xi(t)$ and $\chi_4(t)$ are
well fitted by power law divergencies. Remarkably, size and
lengthscale are related by a robust power law, $\chi_4(\tau_{\alpha})
\approx \xi^{d-p}(\tau_{\alpha})$ with an exponent $d-p\approx 1.6$,
implying that the clusters of slow particles are neither compact nor
string-like. For fixed packing fraction the strength of the dynamical
heterogeneity changes dramatically with the degree of inelasticity,
however the scaling $\chi_4(\tau_{\alpha}) \approx
\xi^{d-p}(\tau_{\alpha})$ and the exponent $d-p$ are universal.  To
obtain these results, and in particular a correlation length as large
as $72$ particle radii, we rely on large scale simulations with
typically $4\times (10^5-10^6)$ particles. We find that in 2d the
spatial fluctuations are much stronger but the relaxation time grows
much more slowly with lengthscale than in 3d.



We consider a bidisperse system of hard disks in 2D, with radii $r_2$
and $r_1$ such that $r_2 \approx 1.43 r_1$. The hard disks interact
via two-body inelastic collisions: The normal component of the
relative velocity of 2 colliding particles is multiplied by a factor
$\varepsilon \le 1$, the coefficient of restitution. In the inelastic
case $\varepsilon < 1$, energy $\propto (1-\varepsilon^2)$ is
dissipated in each collision, and has to be supplied in order to reach
a steady state. Here, we kick the particles randomly, comparably to
the bulk driving in the experiments presented
in~\cite{Abate2007,Keys2007}.  The total injected power is chosen
$\propto (1-\varepsilon^2)$, in order to achieve approximately the
same granular temperature $T_G$ (a measure of the rms velocity
fluctuations), for all $\varepsilon$.  The system presented here is
the same as the one in Ref.~\cite{Gholami2011}, where additional
simulation details can be found.

In this work, we analyze simulations for $\varepsilon = 0.90$ with
packing fractions $0.6\leq\phi\leq 0.805$, and for $\varepsilon=0.70,
0.80, 1.00$ with packing fractions $0.72\leq\phi\leq 0.79$. The system
contains $N_{\text{tot}}=4,000,000$ particles for $0.60 \le \phi \le
0.78$ and $N_{\text{tot}}=360,000$ particles for $0.79 \le \phi \le
0.805$. We measure lengths in units of the radius $r_1$ of the small
disks and choose units of time such that $T_G = 1$. 
%
To analyze the results, we divide the
simulation box, which has total area $L^2_{\text{tot}}$, into
sub-boxes of equal areas $L^2$. The number of particles $N_{\bf r}$ in
each sub-box $B_{\bf r}$ (centered at point ${\bf r}$) fluctuates over
time and between different sub-boxes, but its average $N =
N_{\text{tot}} (L/L_{\text{tot}})^2$ has been kept fixed for each
measurement. For all analysis we select the time window so that the
system is in a steady state. 

To probe the dynamics, we define the single-particle overlap function
$w_i(t_2,t_1) \equiv \theta(a-|{\bf r}_i(t_2)-{\bf r}_i(t_1)|)$, where
$\theta$ is the Heaviside function, $t_1$ and $t_2$ are times such
that $t_2 \ge t_1$, ${\bf r}_i(t)$ is the position of
particle $i$ at time $t$, and $a$ is the cutoff
length. Intuitively, this observable distinguishes between ``slow''
particles, with $w_i=1$, and ``fast'' particles, with $w_i=0$. For
each sub-box $B_{\bf r}$ and for a given time interval between $t_0$
and $t_0+t$ ($t>0$), we also define the sub-box overlap 
$Q_{\bf r}(t;t_0)= \frac{1}{N_{\bf r}}\sum_{i=1}^{N_{\bf r}}
w_i(t_0+t,t_0)$, where the sum runs over the particles present in the
box $B_{\bf r}$ at time $t_0$. $Q_{\bf r}(t;t_0)$ can be
interpreted as the fraction of slow particles in sub-box $B_{\bf r}$
in the time interval $[t_0,t_0+t]$. 

The average dynamics is characterized by the quantity
$\overline{\left\langle Q_ {\bf r}(t;t_0) \right\rangle}$, where
$\langle \cdots \rangle$ denotes an average over sub-boxes, and
$\overline{\cdots}$ denotes an average over initial times $t_0$ at
fixed time difference $t$. This quantity exhibits critical
slowing down as the packing fraction increases~\cite{Avila-PRE}. In
particular, the $\alpha$ relaxation time $\tau_{\alpha}$, defined by
$\overline{\left\langle Q_{\bf r}(\tau_{\alpha};t_0)
  \right\rangle} = 1/e$, is a rapidly increasing function of
$\phi$~\cite{Avila-PRE}. Unless otherwise indicated, the results shown
below are for $\varepsilon=0.9$, $a = 0.6 \, r_1$, and $t =
\tau_{\alpha}$.

\begin{figure}[hb]
  \includegraphics[scale=.377]{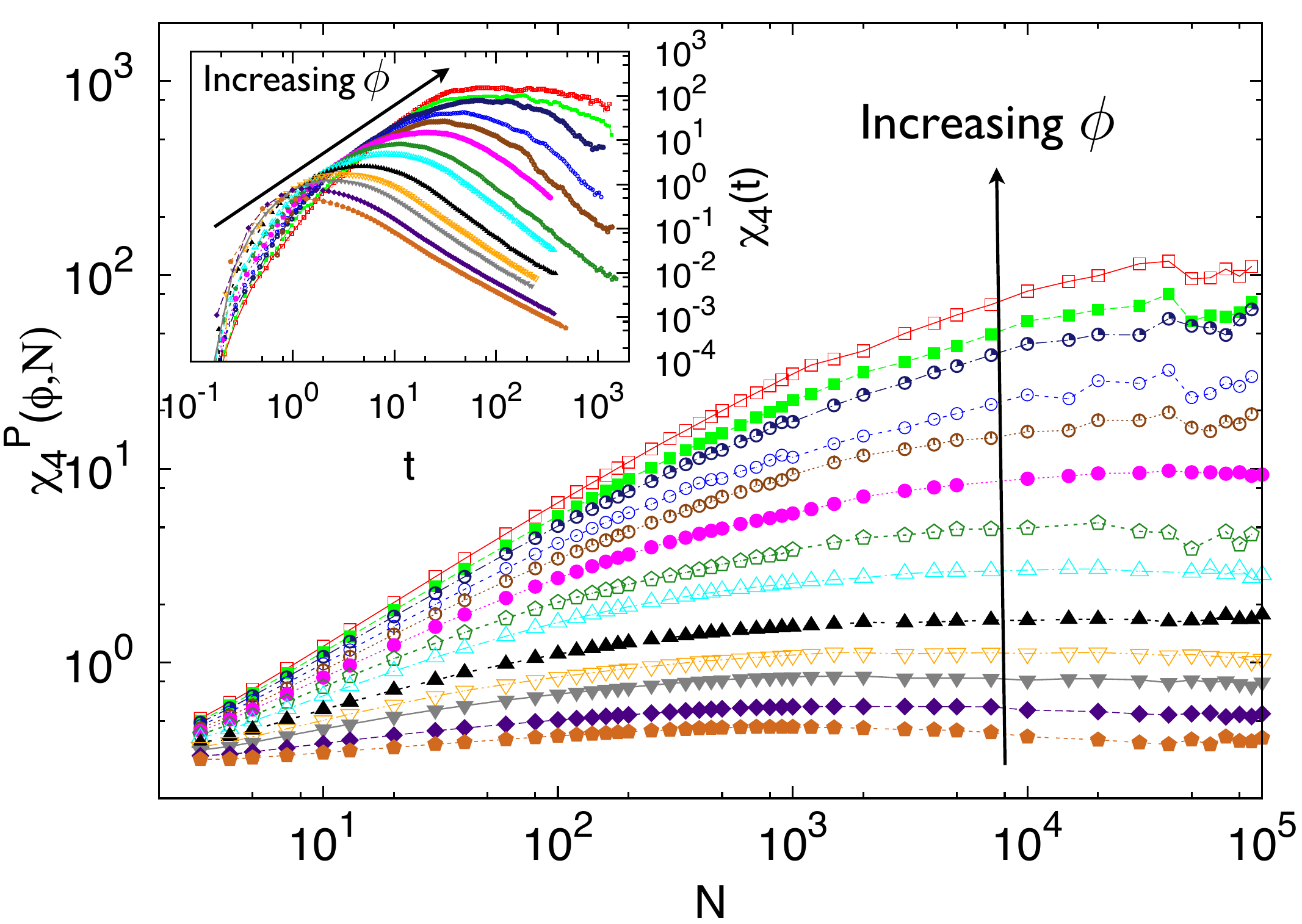}
  \caption{(Color online) Peak value of the dynamic susceptibility,
    $\chi_4^P$, versus $N$ for packing fractions $0.60 \leq \phi \leq
    0.805$.
    {\em Inset:} Dynamic
    susceptibility, $\chi_4(t)$, for the same packing fractions as in
    the main panel, for $N = 10,000$. 
  }
  \label{fig:chi_4-peak}
\end{figure}
To quantify the heterogeneity of the dynamics, we use the dynamic
susceptibility
\begin{equation}
\chi_4(t) = N \overline{ \left[ 
    \left\langle Q_{\bf r}^2(t;t_0) \right\rangle 
    - \left\langle Q_{\bf r}(t;t_0) \right\rangle^2 
    \right] }, 
\label{eq:chi_4}
\end{equation}
which gives a direct measure of the strength of the fluctuations in
the overlap. As a function of time, $\chi_4(t)$ has a maximum
$\chi_4^P$ at time $\tau^*$
. Both the maximum value $\chi_4^P$
and its position $\tau^*$ are increasing functions of the packing
fraction $\phi$ (see Fig.~\ref{fig:chi_4-peak}, inset).
Moreover, as shown in Fig.~\ref{fig:chi_4-peak}, as a function of $N$,
$\chi_4^P$ initially increases and then reaches a plateau. Both the
value of $N$ at which the plateau starts and the plateau value of
$\chi_4^P$ are increasing functions of the packing fraction
$\phi$, consistent with the presence of a
correlation length $\xi$ that controls the finite size scaling
behavior of $\chi_4^P$ and that grows with increasing $\phi$~\cite{Karmakar2009}. 
To minimize finite size effects, in what
follows all results are reported for $N=10,000$ for $\phi \le 0.76$
and $N=40,000$ for $\phi > 0.76$, which are within the plateau region
for all packing fractions considered.

{\it Spatial correlations} of the
dynamical fluctuations
are encoded in the four--point structure factor 
\begin{eqnarray}
\lefteqn{S_4(q,t)/N =} \\[0.5ex]
 & &  \left \{ \overline{ 
    \left[\left\langle W_{\bf r}({\bf q},t;t_0) 
      W_{\bf r}({\bf -q},t;t_0) \right\rangle 
      - \left\langle W_{\bf r}({\bf q},t;t_0) 
      \right\rangle \left\langle 
      W_{\bf r}(-{\bf q},t;t_0)
      \right\rangle  \right]}\right \} , \nonumber
\label{eq:s4-eq}
\end{eqnarray}
where
$W_{\bf r}({\bf q},t;t_0)= \frac{1}{N_{\bf
      r}}\sum_{i=1}^{N_{\bf r}} \exp{[i{\bf q}\cdot {\bf
      r}_i(t_0)]} w_i(t_0+t,t_0)$,
and  $\{ \cdots \}$ denotes an average over wave vectors ${\bf{q}}$ of
fixed magnitude $|{\bf q}| = q$. 
The four-point structure factor and the dynamic susceptibility are
related by $\lim_{q \to 0} S_4(q,t) = \chi_4(t)$~\cite{correction-term}.

\begin{figure}[h]
   \includegraphics[scale=.685]{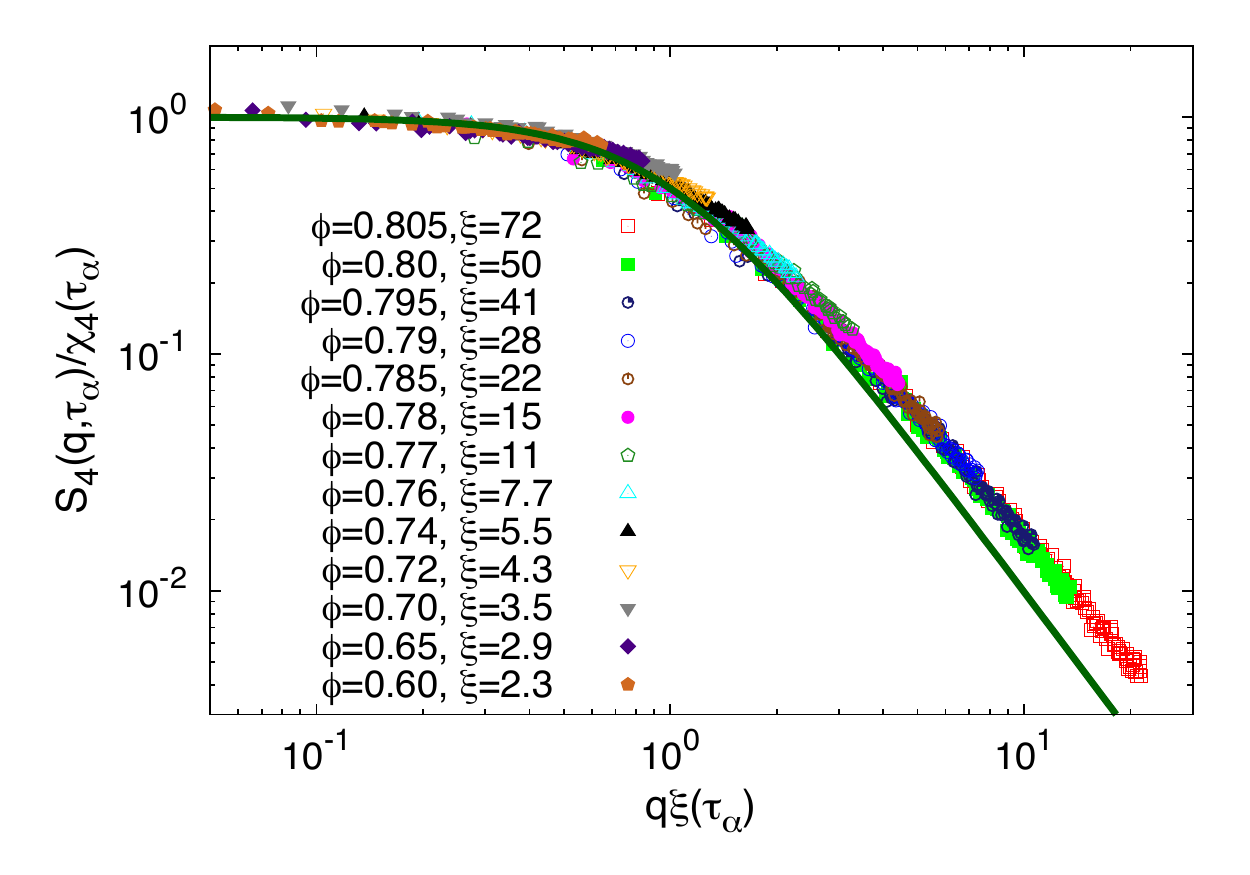}
  \caption{(Color online) Scaling plot of the four--point structure
    factor $S_4(q,\tau_{\alpha})$ for different packing fractions,
    with Ornstein-Zernicke fit (solid line). The correlation lengths
    $\xi$ are shown in the key.  
  }
  \label{fig:S4}
\end{figure} 
As the packing fraction is increased to the point of structural
arrest, we expect long range correlations of the dynamic
heterogeneities as well as scaling of $S_4(q,t)$. 
In Fig.~\ref{fig:S4} we plot
$S_4(q,\tau_{\alpha})/\chi_4(\tau_{\alpha})$ as a function of $q
\xi(\tau_{\alpha})$, and find good collapse between data for
different $\phi$. This shows that all dependence on $\phi$ can be
absorbed into a single lengthscale, the dynamic correlation length
$\xi(t)$ evaluated at $\tau_{\alpha}$. $\xi(t)$ can be extracted
either by collapsing the data in the scaling plot or by fitting
$S_4(q,t)$ to the Ornstein-Zernicke (OZ) form,
$S_4(q,t)=\chi_4(t)/\{1+[{q}\xi(t)]^2\}$.
%
%

As can be seen in Fig.~\ref{fig:S4}, the scaling function is
close to the OZ form for $q \xi(\tau_{\alpha}) \alt 1$, but
starts to differ significantly from it for larger values of $q
\xi(\tau_{\alpha})$. The values of $\xi(\tau_{\alpha})$, reported in
Fig.~\ref{fig:S4}, are obtained by fitting $S_4(q,t)$ to the OZ form in
the range $0 < q < 0.2$. Changing the fitting range or adding a
quartic term to the 
denominator in the fitting function~\cite{Flenner2011} does not
significantly alter the results for $\xi(\tau_{\alpha})$~\cite{Avila-PRE}.

\begin{figure}[t]
  \includegraphics[scale=.37]{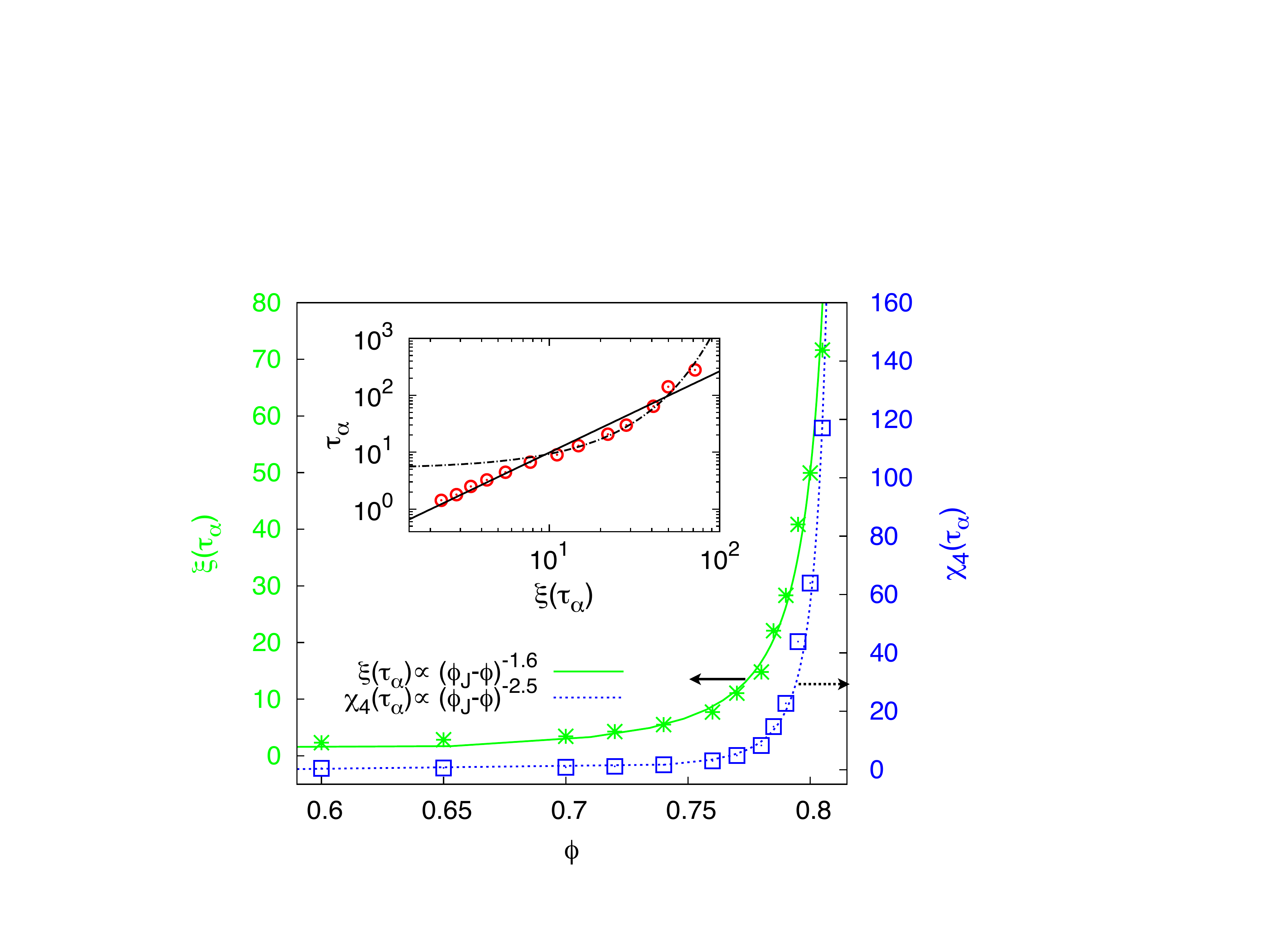}
  \caption{(Color online) {\em Main panel:\/} The dynamic correlation
    length $\xi(\tau_{\alpha})$ and the dynamic susceptibility
    $\chi_4(\tau_{\alpha})$ as functions of the packing fraction
    $\phi$, In each case, a fit to a diverging power law function is
    shown.
    {\em Inset:\/} Relaxation time
    $\tau_{\alpha}$ versus dynamic correlation length
    $\xi(\tau_{\alpha})$. Two fits to the data 
    are attempted: a power law 
    (solid line), and an exponential 
    (dot-dashed line).}
   \label{fig:xi-chi_4}
\end{figure} 

In Fig.~\ref{fig:xi-chi_4}, we show that both $\chi_4(\tau_{\alpha})$
and $\xi(\tau_{\alpha})$ grow rapidly with $\phi$. In fact, both
quantities and also the relaxation time $\tau_{\alpha}$ (not shown)
are well fitted by divergent power law forms $\chi_4(\tau_{\alpha})
\propto (\phi_J-\phi)^{-\gamma_{\chi}}$, $\xi(\tau_{\alpha}) \propto
(\phi_J-\phi)^{-\gamma_{\xi}}$, and $\tau_{\alpha} \propto
(\phi_J-\phi)^{-\gamma_{\tau}}$ with a common location $\phi_J \approx
0.821$ for all three divergences, but different exponents
$\gamma_{\chi} \approx 2.5$, $\gamma_{\xi} \approx 1.6$, and
$\gamma_{\tau} \approx 2.4$~\cite{Avila-PRE}. The latter has been
predicted by mode-coupling~\cite{MCT} ($\gamma_{\tau} \approx 2.5$)
and we expect it to be related to the exponent for the divergence of
the viscosity at jamming~\cite{Olsson,AHB}.

The above results imply a power
law relation between time- and lengthscales: $\tau_{\alpha}
\propto [\xi(\tau_{\alpha})]^{z}$, with a dynamical exponent
$z=\gamma_{\tau}/\gamma_{\xi}$. In the inset of
Fig.~\ref{fig:xi-chi_4} we show $\tau_{\alpha}$ as a function of
$\xi(\tau_{\alpha})$.
A power law (full line) approximately describes the data, 
with a slight deviation at the highest packing fractions, 
and yields an exponent $z = \gamma_{\tau}/\gamma_{\xi} \approx 1.5$.
An alternative
description~\cite{Flenner2011} $\tau_{\alpha}
\propto \exp[k \xi(\tau_{\alpha})]$ (dot-dashed line)
is also shown.
We do not observe the dramatic slowdown of growth of the correlation
volume for very long timescales seen in structural
glasses~\cite{Dalleferrier2007, Harrowell2010}, although we cannot
exclude it happening at lengthscales which exceed the observed
correlation length of 35 particle diameters. This slowdown in glasses
is necessary to avoid unphysically large correlation lengths, when
extrapolated to experimental time scales, but, in a granular fluid,
the timescales are macroscopic and hence time and length scale in the
simulation are comparable to experiment.

\begin{figure}[h]
  \begin{center}
    \includegraphics[scale=.36]{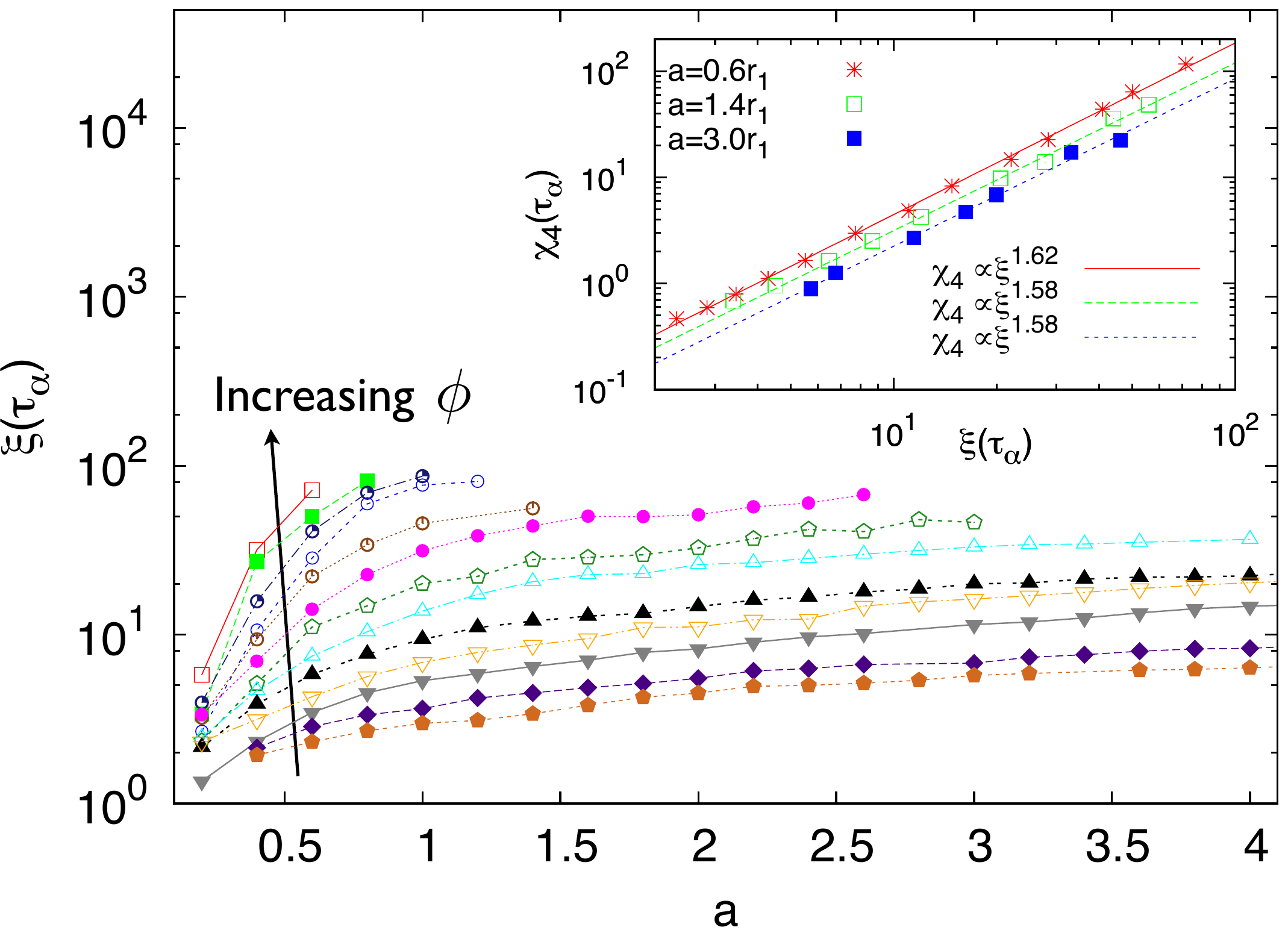}
  \end{center}
  \caption{(Color online) {\em Main panel:\/} $\xi({\tau_{\alpha}})$ versus $a$ for
    $0.60 \leq \phi \leq 0.805$ (from bottom to top). {\em Inset:\/} $\chi_4(\tau_{\alpha})$ versus
    $\xi(\tau_{\alpha})$ for three different choices of the parameter
    $a$. The different lines correspond to the fit
    $\chi_4(\tau_{\alpha})\propto \xi^{d-p}(\tau_{\alpha})$ for each
    $a$.
}
  \label{fig:xi-vs-a}
\end{figure} 

We now examine how the dynamic
susceptibility $\chi_4(\tau_{\alpha})$ and the correlation length
$\xi(\tau_{\alpha})$ depend on $a$. For $a$ within the range $0.2 \,r_1
\le a \le 4.0 \, r_1$, both quantities display the same
behavior~\cite{wider-range-a}. They
grow monotonously with $a$, and three regimes can be identified:
extremely fast growth for $r/a_1 \alt 1$, much slower growth for
$r/a_1 \agt 1$, and a crossover in between. 
Fig.~\ref{fig:xi-vs-a} shows this for the case of
$\xi(\tau_{\alpha})$. We also find that for fixed $a$, the relation
between the two quantities is well fitted by a power law,
$\chi_4(\tau_{\alpha}) \propto \xi^{d-p}(\tau_{\alpha})$, with an
exponent $d-p \approx 1.6$ which is approximately constant as a
function of $a$. In the inset of Fig.~\ref{fig:xi-vs-a} we show this
relation for $a/r_1 = 0.6, 1.4, 3.0$, i.e. for one value of $a$ in
each of the regimes described above.
 
The exponent $d-p$ gives information about the correlated slow
regions. In the most common interpretation, $d-p$ is the fractal
dimension $d_f$ of those regions. The value $d_f \approx 1.6$ differs
from the expected values for compact domains ($d_f = 2$) and for
string-like domains ($d_f = 1$). It has been suggested that
alternatively the correlated regions could be compact, but their sizes
could have a wide
distribution~\cite{Berthier2011,power-law-tail}. However this is not
compatible with the Ornstein-Zernicke form of $S_4(q,t)$, which
implies a fast decay of $G_4(r,t)$ for large distances $r$.
%
We have studied a wide range of values of the cutoff $a$,
which goes from being barely larger than the typical displacement
associated with vibrations of caged particles to being larger than the
displacement required to reach the position of second neighbors to the
original location of the particle. Therefore it is 
remarkable that the exponent $d-p$ is essentially constant over this
whole range of values of $a$.

\begin{figure}[h]
  \begin{center}
   \includegraphics[scale=.29]{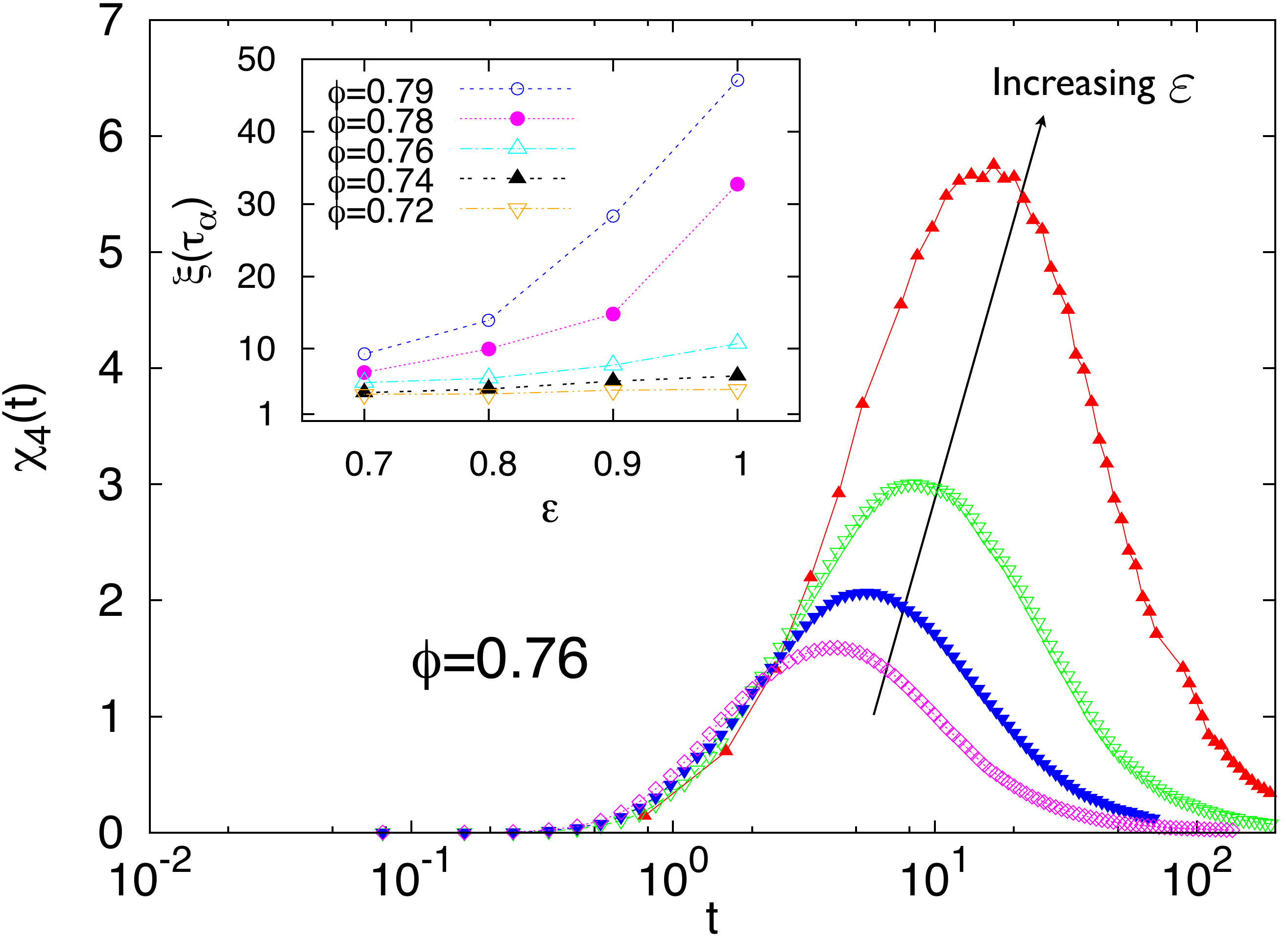}
  \end{center}
  \caption{(Color online) $\chi_4(\tau)$ for coeffcients of restitution 
    $\varepsilon=0.70, 0.80, 0.90, 1.00$ ($\phi=0.76$ fixed). 
    {\em Inset:} Correlation length $\xi(\tau_{\alpha})$ as a
    function of $\varepsilon$, for $0.72 \leq \phi \leq 0.79$.}
  \label{fig:diff-eps}
\end{figure}

We now turn to the analysis of the effects of dissipation by
comparing results for different values of the coefficient of
restitution $\varepsilon$. In Fig.~\ref{fig:diff-eps} we show the
dynamic susceptibility $\chi_4(t)$ for $\phi=0.76$ and
$\varepsilon=0.70, 0.80, 0.90$ and $1.00$ (elastic). 
As $\varepsilon$ grows, the height of the peak of $\chi_4(t)$ increases
and the peak shifts to longer times.  In the inset we show that
$\xi(\tau_{\alpha})$ also grows as a function of
$\varepsilon$ and that this growth is stronger for higher packing fractions.%
\begin{figure}[h]
  \begin{center}
\includegraphics[scale=.66]{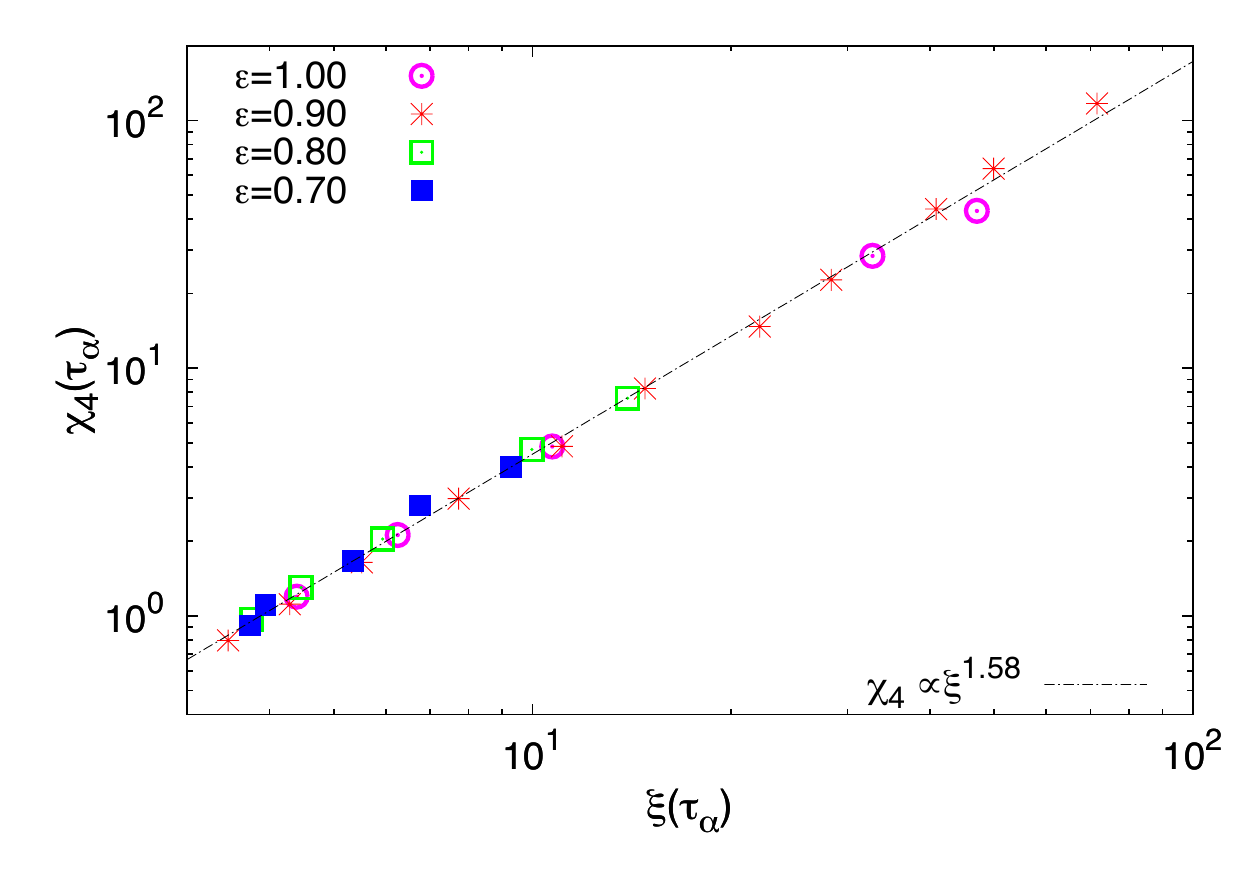}
  \end{center}
  \caption{(Color online) $\chi_4(\tau_\alpha)$ against
    $\xi(\tau_{\alpha})$ for $0.70 \le \varepsilon \le 1.00$ and
    $0.72 \le \phi \le 0.79$. All data are fitted to
    $\chi_4(\tau_{\alpha})\propto \xi^{d-p}(\tau_{\alpha})$, with $d-p
    \approx 1.59$ (dot-dashed line).}
  \label{fig:diff-eps-2}
\end{figure} 
Both results are compatible with an $\varepsilon$--dependent critical
density $\phi_J(\varepsilon)$ as predicted in \cite{MCT}. 
Such a shift in the critical density drops out if we plot the
relation between $\chi_4(\tau_{\alpha})$ and ${\xi(\tau_{\alpha})}$ as
is done in Fig.~\ref{fig:diff-eps-2} for $\varepsilon=0.70, 0.80, 0.90$
and $1.00$. We find that a single power law $\chi_4(\tau_{\alpha})
\propto \xi^{d-p}(\tau_{\alpha})$, with $d-p \approx 1.6$, provides a
good fit for the data corresponding to all values of $\varepsilon$. In
fact, attempting separate fits for each $\varepsilon$ leads to
obtaining exponents that are equal to each other within 
error bars.

In summary, we studied dynamical heterogeneity in a 2d driven granular
fluid in the range of packing fractions $0.6 \leq \phi \leq
0.805$. The 4-point dynamic structure factor was shown to obey
scaling,
$S_4(q,\tau_{\alpha})/\chi_4(\tau_{\alpha})=s(q\xi(\tau_{\alpha}))$,
where the scaling function is well fitted by the Ornstein-Zernicke
form for small argument. This allowed us to determine the dynamic
susceptibility $\chi_4(\tau_{\alpha})$ and the correlation length
$\xi(\tau_{\alpha})$ independently. Both were shown to grow
dramatically with the packing fraction $\phi$ and can be well fitted
by divergent power laws within the range of packing fractions accessible to our simulations. For restitution coefficients $0.7 \le
\varepsilon \le 1.0$, and a wide range of cutoffs $0.6 \le a/r_1 \le
3.0$, we found a robust scaling $\chi_4(\tau_{\alpha}) \propto
\xi^{d-p}(\tau_{\alpha})$, with $d-p \approx 1.6$, implying that the
correlated regions are neither string-like nor compact. We conclude
that the observed scaling of dynamical heterogeneities is remarkably
universal with respect to dissipation and much stronger in 2d than in
3d.


\section{Acknowledgments}

H.E.C. thanks E.~Flenner, and G.~Szamel for discussions. This work
was supported in part by DFG under grants SFB 602 and FOR 1394, by DOE
under grant DE-FG02-06ER46300, by NSF under grants PHY99-07949 and
PHY05-51164, and by Ohio University. K.E.A. acknowledges the CMSS
program at Ohio University for partial support. We thank I.~Gholami
and T.~Kranz for help with the numerical simulations.


\begin{thebibliography}{10}
\bibitem{power-law-tail} 
%
One way in which $d-p < 2$ could be obtained, even if the cluster
fractal dimension $d_f$ is very close to $2$, is if the distribution
of cluster linear sizes $\hat{\xi}$ has a power law tail with a cutoff
$\rho(\hat{\xi}) \propto \theta({\hat{\xi}}_{\rm max} - \hat{\xi})
{\hat{\xi}}^{-\alpha}$, with $\alpha = 3 -
\frac{2}{p}(2-d_f)$~\cite{Avila-PRE}. In particular, $d-p = 1.6$ would
correspond to $\alpha = 3 - 5(2-d_f)$. This is based on the assumption
that the correlation length $\xi$ that is extracted from the small $q$
limit of $S_4(q,t)$ is an estimate of the rms value 
$\xi = \langle {\hat{\xi}}^2 \rangle^{1/2}$.
\end{thebibliography}
\end{document}